\documentclass[12pt]{article}         
\usepackage{graphicx}         
         
\begin{document}         
\hfill{FTUV-01-0216}  
  
\hfill{IFIC/01-05}  
  
\hfill{TPR-01-01}  
   
\vspace{1.cm}   
        
\begin{center}         
{\Large\bf         
Wandzura-Wilczek approximation          
from generalized rotational invariance}         
         
\vspace{.5cm}         
{\large\rm{         
I.V.~Anikin$^{\,a\, ,\, b \, , \,c}$ ,         
O.V.~Teryaev$^{\,a\, , \,c}$}}         
         
\vspace{.5cm}         
{\it $^{a}$ Institut f\"{u}r Theoretische Physik,         
Universit\"{a}t Regensburg,         
D-93040 Regensburg, Germany}  \\          
{\it $^{b}$Departament de F\`{i}sica Te\`{o}rica         
and Institut de F\`{i}sica Corpuscular,         
Universitat de Val\`{e}ncia-CSIC         
E-46100 Burjassot (Val\`{e}ncia), Spain}\\         
{\it $^{c}$ Bogoliubov Laboratory of Theoretical Physics,         
Joint Institute for Nuclear Research,         
141980 Dubna, Russia }         
\end{center}         
                        
\begin{abstract}         
We present the alternative way of derivation of          
the Wandzura-Wilczek relations          
between the kinematical twist-3 and twist-2 functions, parameterizing          
hadronic matrix element in two-photon processes          
$\gamma^{\star}\pi\to         
\gamma\pi$ and $\gamma^{\star}\gamma\to\pi\pi$.         
The new equations, providing the independence of          
the physical cross-sections on the choice of the light-cone direction,         
are suggested and explored.           
The amplitude of $\gamma^{\star}\gamma\to\pi\pi$ up to         
genuine twist-3 accuracy is found.         
\end{abstract}         
         
\vspace{1cm}         
\noindent         
{\it Introduction.} The investigation of the          
two-photon amplitudes of Deeply Virtual Compton Scattering (DVCS)         
or virtual photon production of meson pairs,         
up to $1/Q$ corrections (twist-3 terms)         
provides one of the interesting lines of inquiry.            
Indeed, these corrections ensure the electromagnetic         
gauge invariance of the DVCS amplitude with the sizeable          
transverse components of the momentum transfer. This problem was         
discussed  in the pioneering paper \cite{Ji}         
and the phenomenological         
solution was presented in         
\cite{G&V}.         
The field-theoretical treatment was suggested in          
\cite{PT98,AN00} and developed in \cite{Pitt,B&M,R&W}.          
From the other side, the study of these corrections provides             
the possibility to consider the delicate effects of factorization          
\cite{Kivel,R&W}.          
        
It was shown \cite{B&M}, \cite{Kivel} that,          
if the dynamical (genuine)          
high twist terms due to the physical transverse components          
of gluonic fields are neglected,           
the parameterizing functions (the skewed, or generalized,      
parton distribution)         
of kinematical         
twist-3 can be related to the parameterizing functions of twist-2         
by means of the Wandzura-Wilczek (WW) relations.              
Their general structure            
was investigated, and it was demonstrated that within          
the WW approximation the skewed twist-3 distribution          
functions have the discontinuities at $x=\pm\xi$,         
which fortunately do not lead to the observable consequences         
at the level of twist-3 accuracy. However, the general relevance          
of these effects (and, in particular, their possible appearance in the         
genuine twist case, for both DVCS and meson pair production channels)    
is still obscure. To some extent, this may be due to the      
lack of the physical interpretation of the WW approximation,         
essentially based on the involved treatment of the          
light-cone operators with total derivatives    
in the coordinate representation.         
   
In this note we present the alternative derivation of the WW relations         
for two-photon processes based on the Ellis-Furmanski-Petronzio (EFP)     
\cite{EFP}         
factorization scheme. In the framework of EFP scheme, the separation          
of the kinematical and          
dynamical twist-3 effects is rather physical and amounts to the          
taking into account the terms linear in the intrinsic transverse momentum    
of partons in    
hadrons \footnote{This method in the case of DIS was developed          
by one of the authors a few years ago \cite{OT95}}.         
As a result, we present the decomposition          
of parameterizing functions of twist-3 to          
the WW-type contribution and contribution         
arising from the three-particle correlators.         
We apply this method for the         
amplitude           
$\gamma^{\star}\gamma\to\pi\pi$. As a result, it is calculated          
keeping both kinematical and genuine twist-3 corrections.        
We found the analogs of the mentioned singularities         
and show that the physical amplitude at twist-3 level is        
not affected by them.

\noindent         
{\it DVCS amplitude of pions in WW approximation.}          
As in \cite{AN00}, we adhere the         
EFP factorization scheme.         
Neglecting the dynamical twist-3 contribution,          
which amounts to the keeping of the operators with transverse derivatives         
alone (c.f. \cite{OT95}),         
being just the Fourier transform of the mentioned intrinsic          
transverse momentum,          
we are able to write the DVCS amplitude for the pion target          
in the following form:         
\begin{eqnarray}         
\label{1.1}         
&&T_{\mu\nu}^{(a)}+T_{\mu\nu}^{(b)}=         
\int dx{\rm tr}\biggl\{ E_{\mu\nu}(x P) \Gamma (x) \biggr\}         
+         
\nonumber\\         
&&\int dx_1 dx_2{\rm tr}\biggl\{         
E_{\mu\rho\nu}(x_1 P, x_2 P) \omega_{\rho\rho^{\prime}}         
\Gamma_{\rho^{\prime}} (x_1, x_2) \biggr\}         
\end{eqnarray}         
where $\omega_{\rho\rho^{\prime}}=\delta_{\rho\rho^{\prime}}-         
n_{\rho^{\prime}} P_{\rho}$,         
and         
\begin{eqnarray}         
\label{2.1}         
&&\Gamma_{\alpha\beta}(x)=-\int d\lambda \,         
e^{ i( x+\xi)\lambda }         
\langle p^{\prime} |         
\psi_{\alpha}(\lambda n) \bar\psi_{\beta}(0)| p \rangle ,         
\nonumber\\         
&&\Gamma^{\rho^{\prime}}_{\alpha\beta} (x_1, x_2)=         
-\int d\lambda_1 d\lambda_2 \,         
e^{ i( x_1+\xi) \lambda_1         
+i(x_2-x_1) \lambda_2 }         
\nonumber\\         
&&\langle p^{\prime} |         
\psi_{\alpha}(\lambda_1 n)         
i\stackrel{ \leftrightarrow }         
{ \partial^{\rho^{\prime}} }(\lambda_2 n)         
\bar\psi_{\beta}(0)         
| p \rangle ,         
\end{eqnarray}         
where $p^{\prime}=P+\Delta/2$,          
$p=P-\Delta/2$, $\Delta= q-q^{\prime}=-2\xi P+\Delta_T$ and         
$p(p^\prime)$, $q(q^\prime)$ are the initial (final) momenta of          
pion and photon.         
The $T_{\mu\nu}^{(a)}$ and $T_{\mu\nu}^{(b)}$ denotes         
the contributions from the twist-2  and twist-3 amplitudes         
(c.f. \cite{AN00}).          
         
\noindent         
{\it Parameterization of hadronic matrix elements.}          
In terms of the light-cone basis, which consists of the vectors $P$, $n$         
($P^2=n^2=0$ and $P \cdot n=1$)         
and $\Delta_T$, the parametrization of the          
vector and axial hadronic matrix elements (other structures do not contribute         
for massless quarks) can be introduced, as in \cite{AN00}.         
Namely         
\footnote{here $         
\varepsilon_{\rho\Delta^{T}Pn} \equiv \varepsilon_{\rho \alpha \beta \gamma}         
\Delta^{T \alpha} P^\beta n^\gamma$.},          
\begin{eqnarray}         
\label{par1}         
&& \langle p^{\prime} |        
\bar\psi(0)\gamma_{\mu} \psi(\lambda n)         
| p \rangle         
=         
\int\limits_{-1}^{1}dx e^{ -i( x+\xi)\lambda }         
\Biggl\{H_1(x)P_{\mu}+H_3(x)\Delta^{T}_{\mu}         
\Biggr\},         
\\         
\label{par11}         
&&\langle p^{\prime} |         
\bar\psi(0)\gamma_{\mu}         
\stackrel{\leftrightarrow}{i\partial^T_{\rho}}          
 \psi(\lambda n)        
| p \rangle         
=         
\int\limits_{-1}^{1}dx e^{ -i( x+\xi)\lambda}         
H_1^{T}(x)P_{\mu}\Delta^T_{\rho},         
\\         
\label{par1.1}         
&& \langle p^{\prime} |          
\bar\psi(0)\gamma_5\gamma_{\mu} \psi(\lambda n)        
| p \rangle         
=         
\int\limits_{-1}^{1}dx e^{ -i( x+\xi)\lambda }         
iH_A(x)\varepsilon_{\mu\Delta^{T}Pn},         
\\         
\label{par1.11}         
&&\langle p^{\prime} |         
\bar\psi(0)\gamma_5\gamma_{\mu}         
\stackrel{\leftrightarrow}{i\partial^T_{\rho}}          
\psi(\lambda n)         
| p \rangle         
=         
\int\limits_{-1}^{1}dx e^{ -i( x+\xi)\lambda}         
iH_A^{T} (x) P_{\mu}\varepsilon_{\rho\Delta^{T}Pn},         
\end{eqnarray}         
for the quark correlators, and ,         
\begin{eqnarray}         
\label{par1.2}         
&& \langle p^{\prime} |         
\bar\psi(0)\gamma_{\mu}g A_{\rho}^T(\lambda_2 n)          
\psi(\lambda_1 n)         
| p \rangle         
=         
\nonumber\\         
&&\int\limits_{-1}^{1}dx_1dx_2 e^{ -i( x_1+\xi)\lambda_1 -         
i(x_2-x_1)\lambda_2}         
B(x_1,x_2)P_{\mu}\Delta^{T}_{\rho},         
\\         
\label{par1.3}         
&& \langle p^{\prime} |         
\bar\psi(0)\gamma_5\gamma_{\mu} g A_{\rho}^T(\lambda_2 n)          
\psi(\lambda_1 n)         
| p \rangle         
=         
\nonumber\\         
&&\int\limits_{-1}^{1}dx_1dx_2 e^{ -i( x_1+\xi)\lambda_1 -         
i(x_2-x_1)\lambda_2}         
i D(x_1,x_2)P_{\mu}\varepsilon_{\rho\Delta^{T}Pn},         
\end{eqnarray}         
for the quark-gluon correlators.        
Here we see the contributions due to the quark intrinsic        
transverse momentum and transverse component of the        
gluon field, respectively.       
In the WW approximation        
one is keeping only (\ref{par11}, \ref{par1.11}), neglecting      
(\ref{par1.2}, \ref{par1.3}),     
and in this sense it is already present in \cite{AN00}.       
The remaining ingredient is to express these functions        
in terms of the leading twist one, which can be done         
using the specific version of the rotational invariance.           
         
\noindent         
{\it The $n$-independence of DVCS amplitude.}    
Using the parameterization          
(\ref{par1})-(\ref{par1.11}) we write the expression for          
(\ref{1.1}) in the form         
\begin{eqnarray}         
\label{ampl1}         
&&T_{\mu\nu}^{(a)}+T_{\mu\nu}^{(b)}=         
\int dx {\rm tr}\biggl\{ E_{\mu\nu}(x) \hat P\biggr\}H_1(x,\xi) +         
\nonumber\\         
&&\int dx{\rm tr}\biggl\{ E_{\mu\nu}(x) \hat \Delta^T \biggr\}H_3(x,\xi) +         
\nonumber\\         
&&\int dx {\rm tr}\biggl\{ E_{\mu\nu}(x) \gamma^{\rho}\gamma_5         
\biggr\} i H_A(x,\xi) \epsilon_{\rho\Delta^T P n} +         
\nonumber\\         
&&\int dx_1 dx_2 {\rm tr}\biggl\{         
E_{\mu\rho\nu}(x_1,x_2) \hat P\biggr\}          
H_1^T(x_2,\xi)\delta(x_1-x_2)\Delta^T_{\rho} +         
\nonumber\\         
&&\int dx_1 dx_2 {\rm tr}\biggl\{         
E_{\mu\rho\nu}(x_1,x_2) \hat P\gamma_5 \biggr\}          
i H_A^T(x_2,\xi)\delta(x_1-x_2)         
\epsilon_{\rho\Delta^T P n}.         
\end{eqnarray}         
The gauge invariance requires that the physical amplitudes        
are independent on          
the gauge-fixing  parameter which, in our case, is the      
light-cone vector $n$.         
As it is also used to define the light-cone direction, the same is      
required          
by Lorentz invariance, which should be implemented keeping the constraints      
($n^2=0 ,P \cdot n=1$).     
We are therefore dealing here         
with a particular subgroup of the Lorentz group.      
The relevant symmetry may be called {\it generalized}      
rotational invariance, to distingush with the standard      
rotational invariance in a particular rest frame,      
and to stress its specific form in the case of generalized parton      
distribution, to be discussed below.       
One should recall that the Burkhardt-Cottingham sum rule, known to be      
intimately related to rotational invariance, follows from the WW      
approximation for polaritzed DIS, and may be derived from      
$n$ independence \cite{Efr84,OT95}.         
Note that it is rather Lorentz than gauge invariance which is crucial     
here, as one may consider the covariant gauge, where role of $n$ as a      
gauge-fixing vector is absent \footnote{The new terms, proportional to       
$A \cdot n$ are absorbed to the standard P-exponent    
of the leading twist      
operators in this case \cite{SQ}}.     
Formally, this invariance can be written as         
\begin{eqnarray}         
\label{n-ind_con1}         
T_{\mu\nu} [n+\delta n]=T_{\mu\nu} [n],     
\end{eqnarray}     
or, equivalently,      
\begin{eqnarray}         
\label{n-ind_con}         
\frac{d}{dn_{\beta}}\biggl\{         
T_{\mu\nu}^{(a)}+T_{\mu\nu}^{(b)}         
\biggr\}= 0,         
\end{eqnarray}     
Note the essential difference with the (unpolarized)     
DIS case \cite{EFP}, where dependence          
on $n$ was coming only through the explicit dependence of the          
appearing tensor structures. Here, it comes also through the          
$\xi$ dependence of the matrix elements, so that          
\begin{eqnarray}         
\label{n-tot}         
\frac{d}{dn_{\beta}}=          
\frac{\partial}{\partial n_{\beta}}         
- \frac {\Delta^\beta}{2} \cdot \frac{\partial}{\partial \xi}.         
\end{eqnarray}         
The second term does not appear also      
in the polarized DIS case \cite{Efr84} (where the role          
of $\Delta$ is played by the polarization vector $S$), as the dependence          
of parton distributions and correlations on $(S\cdot n)$, being the analog      
of          
$\xi$, is absent due to the general property of the          
linearity of density matrix.      
It is this extra term, specific to the case of generalized     
parton distributions, which is one of the reasons to use the term     
'generalized' for this particular version of rotational invariance.        
        
Note that the relation (\ref{n-ind_con}) is in some sense analogous to the          
renormalization group equation governing the dependence of hadronic matrix          
elements on the factorization scale. Indeed, the factorization scale $\mu$          
is the unphysical auxiliary           
parameter required to write the physical amplitude         
in the factorized form. Analogously, $n$ is          
the auxiliary vector, required to specify the infinite-momentum frame.          
The dependence of the observable quantities on both these parameters,         
$\mu$ and  $n$, must vanish.         
         
Using  the Ward identity, the computation of the derivative in          
(\ref{n-ind_con}) leads to the following equation:          
\begin{eqnarray}         
\label{n-ind_con2}         
&&\frac{d}{dn_{\beta}}\biggl\{         
T_{\mu\nu}^{(a)}+T_{\mu\nu}^{(b)}         
\biggr\}=         
-\frac{1}{2}\int dx {\rm tr}\biggl\{ E_{\mu\nu}(x) \hat P\biggr\}         
\frac{\partial H_1(x,\xi)}{\partial \xi}\Delta_{\beta} -         
\nonumber\\         
&&\int dx{\rm tr}\biggl\{ E_{\mu\nu}(x) \hat P \biggr\}         
\Delta_{\beta} H_3(x,\xi) +         
\int dx {\rm tr}\biggl\{ E_{\mu\nu}(x) \gamma^{\rho}\gamma_5         
\biggr\} iH_A(x,\xi) \epsilon_{\rho\Delta^T P \beta} -         
\nonumber\\         
&&\int dx_1 dx_2 {\rm tr}\Biggl\{         
\biggl( E_{\mu\nu}(x_1) - E_{\mu\nu}(x_2) \biggr)          
\hat P\Biggr\}          
\frac{H_1^T(x_2,\xi)\delta(x_1-x_2)}{x_1-x_2}\Delta_{\beta} +         
\nonumber\\         
&&\int dx_1 dx_2 {\rm tr}\Biggl\{         
\biggl( E_{\mu\nu}(x_1) - E_{\mu\nu}(x_2) \biggr)          
\gamma_\rho \gamma_5 \Biggr\}          
\frac{i H_A^T(x_2,\xi)\delta(x_1-x_2)}{x_1-x_2}         
\epsilon_{\rho\Delta^T P \beta} = 0.          
\nonumber\\         
\end{eqnarray}          
 Note that we actually did not use the specific form of the hard         
scattering amplitude, so this relation should be valid for          
any hard process.          
Requiring the validity of this equation for arbitrary  $E_{\mu\nu}$,         
we can easily derive the          
$n$-independence condition in the more simple form:         
\begin{eqnarray}         
\label{n-indep_con3}         
&&\frac{\partial H_1^T(x,\xi)}{\partial x}=         
\frac{1}{2}\frac{\partial H_1(x,\xi)}{\partial\xi}+H_3(x,\xi),         
\nonumber\\         
&&\frac{\partial H_A^T(x,\xi)}{\partial x}=H_A(x,\xi).         
\end{eqnarray}         
         
\noindent         
{\it QCD equations of motion for matrix elements.}         
Following \cite{AN00}, we write the QCD equations of motion          
\begin{eqnarray}         
\label{me2.1.1}         
&&\int\limits_{-1}^{1} dy \left( B^{(A)}(x,y)-D^{(S)}(x,y)         
- \delta(x-y)H_A^{T}(y) \right) =         
\nonumber\\         
&&-\xi H_3(x) - \frac{1}{2}H_1(x) - x H_A(x) ,         
\nonumber\\         
&&\int\limits_{-1}^{1} dy \left(B^{(S)}(x,y) + \delta(x-y)H_1^{T}(y)         
-D^{(A)}(x,y)\right) =         
\nonumber\\         
&&xH_3(x)+\xi H_A(x),         
\end{eqnarray}         
where symmetrical and anti-symmetrical functions are defined as,         
\begin{eqnarray}         
\label{saf}         
&& B^{(S,A)}(x,y)=\frac{1}{2}         
\left( B(x,y) \pm B(y,x) \right).         
\end{eqnarray}         
Within the WW approximation they      
can been rewritten as          
\begin{eqnarray}         
\label{em_ww}         
&&H_A^T(x,\xi)=\xi H_3(x,\xi)+\frac{1}{2}H_1(x,\xi)+xH_A(x,\xi),         
\nonumber\\         
&&H_1^T(x,\xi)=x H_3(x,\xi)+\xi H_A(x,\xi).         
\end{eqnarray}         
Note that these equations of course contain the so-called     
total derivatives, crucial for the derivation of WW relations      
\cite{B&M,Kivel,R&W}.      
         
\noindent         
{\it The WW relations for DVCS process.}   
Calculating the derivative of          
(\ref{em_ww}) with respect to        
$x$ and          
using (\ref{n-indep_con3}), we obtain the system of          
differential equations:         
\begin{eqnarray}         
\label{diff_sys}         
&&\frac{\partial H_+(x,\xi)}{\partial x} =         
\frac{1}{2}\frac{\partial_- H_1(x,\xi)}{x+\xi} ,         
\nonumber\\         
&&\frac{\partial H_-(x,\xi)}{\partial x} =         
\frac{1}{2}\frac{\partial_+ H_1(x,\xi)}{x-\xi} ,         
\end{eqnarray}         
where we introduced the notations:         
\begin{eqnarray}         
\label{def_pm}         
&&\partial_{\pm} H_1(x,\xi)=\frac{\partial H_1(x,\xi)}{\partial\xi}         
\pm         
\frac{\partial H_1(x,\xi)}{\partial x},         
\nonumber\\         
&&H_{\pm}(x,\xi)=H_3(x,\xi)\pm H_A(x,\xi).         
\end{eqnarray}         
Taking into account the boundary conditions         
\begin{eqnarray}         
\label{b_con}         
H_{\pm}(1,\xi)=H_{\pm}(-1,\xi)=0,         
\end{eqnarray}           
solution of the system (\ref{diff_sys}) can          
be easily found:          
\begin{eqnarray}         
\label{sol}         
&&H_+^{WW}(x,\xi)=-\frac{1}{2}\Theta(x > -\xi)\int\limits_{x}^{1}dy         
\frac{\partial_- H_1(y,\xi)}{y+\xi} +         
\frac{1}{2}\Theta(x < -\xi)\int\limits_{-1}^{x}dy         
\frac{\partial_- H_1(y,\xi)}{y+\xi} ,         
\nonumber\\         
&&H_-^{WW}(x,\xi)=-\frac{1}{2}\Theta(x > \xi)\int\limits_{x}^{1}dy         
\frac{\partial_+ H_1(y,\xi)}{y-\xi} +         
\frac{1}{2}\Theta(x < \xi)\int\limits_{-1}^{x}dy         
\frac{\partial_+ H_1(y,\xi)}{y-\xi} .         
\nonumber\\   
\end{eqnarray}      
The theta-functions appearance is due to the fact, that we      
are using the different boundary conditions below and above the      
singular points $x=\pm \xi$.          
Expressing the          
$H_{3(A)}^{WW}-$functions by making use of (\ref{def_pm}),         
we have:          
\begin{eqnarray}         
\label{sol_2}         
&&H_3^{WW}(x,\xi)=-\frac{1}{4}\Theta(x > -\xi)   
\int\limits_{x}^{1}dy         
\frac{\partial_- H_1(y,\xi)}{y+\xi}          
-\frac{1}{4}\Theta(x > \xi)   
\int\limits_{x}^{1}dy         
\frac{\partial_+ H_1(y,\xi)}{y-\xi} +         
\nonumber\\         
&&\frac{1}{4}\Theta(x < -\xi)\int\limits_{-1}^{x}dy         
\frac{\partial_- H_1(y,\xi)}{y+\xi} +         
\frac{1}{4}\Theta(x < \xi)\int\limits_{-1}^{x}dy   
\frac{\partial_+ H_1(y,\xi)}{y-\xi} ,        
\nonumber\\         
&&H_A^{WW}(x,\xi)=-\frac{1}{4}\Theta(x > -\xi)\int\limits_{x}^{1}dy         
\frac{\partial_- H_1(y,\xi)}{y+\xi} +         
\frac{1}{4}\Theta(x > \xi)\int\limits_{x}^{1}dy    
\frac{\partial_+ H_1(y,\xi)}{y-\xi}         
 +         
\nonumber\\         
&&\frac{1}{4}\Theta(x < -\xi)\int\limits_{-1}^{x}dy               
\frac{\partial_- H_1(y,\xi)}{y+\xi} -    
\frac{1}{4}\Theta(x < \xi)\int\limits_{-1}^{x}dy         
\frac{\partial_+ H_1(y,\xi)}{y-\xi} .         
\nonumber\\   
\end{eqnarray}            
Note that a given solution completely coincide with the          
expressions presented in ref. \cite{Kivel}, although our          
representation for $H_A^{WW}$ looks a bit more simple.         
   
Using these results, we can now decompose the $H_{3(A)}-$functions         
to the WW type part and genuine part arising from the three-particles          
correlators. Using (\ref{me2.1.1})         
and (\ref{sol}), we obtain the following           
\begin{eqnarray}         
\label{full_h_3}         
H_3(x,\xi) &=& H_3^{WW}(x,\xi) +          
\frac{1}{2(x-\xi)} \int\limits_{-1}^{1}dy         
\biggl( B(x,y,\xi) - D(x,y,\xi)         
\biggr) +         
\nonumber\\         
&&\frac{1}{2(x+\xi)} \int\limits_{-1}^{1}dy         
\biggl( B(x,y,-\xi) - D(x,y,-\xi)         
\biggr)         
\end{eqnarray}         
and          
\begin{eqnarray}         
\label{full_h_a}         
H_A(x,\xi) &=& H_A^{WW}(x,\xi) -          
\frac{1}{2(x-\xi)} \int\limits_{-1}^{1}dy         
\biggl( B(x,y,\xi) - D(x,y,\xi)         
\biggr) +         
\nonumber\\         
&&\frac{1}{2(x+\xi)} \int\limits_{-1}^{1}dy         
\biggl( B(x,y,-\xi) - D(x,y,-\xi)         
\biggr).         
\end{eqnarray}          
Here we take into account the properties of $B-$ and $D-$ functions         
following from the $T$-invariance, {\it i.e.}         
\begin{eqnarray}         
\label{TG}         
B(x,y,\xi) = B (y,x,-\xi), \,\,         
D(x,y,\xi) = - D (y,x,-\xi).         
\end{eqnarray}         
          
We can see that the functions $H_+$ and $H_-$         
(both in WW approximation and beyond) possess the          
singularity at points $-\xi$ and $+\xi$, respectively.         
Consequently, as it was mentioned in \cite{Kivel} and confirmed in         
\cite{R&W}, the functions $H_3$ and $H_A$ have  finite jumps        
in these points. One can add here, that        
if the function $H_1(\xi,\xi)$ is finite,         
this singularity is in fact logarithmical, and the mentioned jump        
means nothing more than the finite difference between two         
logarithmically growing values $H_{3,A}(x \pm \epsilon)$.            
        
From the other side, in the         
DVCS twist-3 amplitude we are dealing with          
the following combinations of $H_+$ and $H_-$          
\cite{AN00},\cite{Kivel} :         
\begin{eqnarray}         
\label{com1}         
\Delta_{\nu}^T\Biggl( 3\xi P_{\mu}+Q_{\mu}         
\Biggr)         
\Biggl( H_3(x,\xi)+\frac{\xi}{x}H_A(x,\xi)         
\Biggr),         
\end{eqnarray}          
which reduce to the $H_+-$structure at $x=\xi$          
and  $H_--$structure at $x=-\xi$,          
and         
\begin{eqnarray}         
\label{com2}         
\Delta_{\mu}^T\Biggl( \xi P_{\nu}+Q_{\nu}         
\Biggr)         
\Biggl( H_3(x,\xi)-\frac{\xi}{x}H_A(x,\xi)         
\Biggr),         
\end{eqnarray}          
which reduces to the $H_- -$structure at $x=\xi$        
and  $H_+ -$structure at $x=-\xi$.         
         
In (\ref{com1}) and (\ref{com2}) the indices $\mu$ and $\nu$         
belong the virtual and real photons, respectively.         
Therefore the DVCS amplitude with the longitudinal polarization          
of virtual photon (\ref{com1}) is free from the divergences.        
At the same time, in the          
case of the transverse polarized virtual photon (\ref{com2})         
the contribution from twist-3 amplitude to          
the observables is suppressed as $1/Q^2$ order after the contraction        
with the polarization vector           
\cite{R&W,Kivel}.

\noindent         
{\it $\gamma\gamma^*\to\pi\pi$ process.}   
We are now passing to the consideration of         
$\gamma(q^\prime)\gamma^*(q)\to\pi(p^\prime)\pi(p)$ process,        
adopting the new notations for        
$P=p+p^{\prime}=q+q^{\prime}$ and           
$\Delta= p^{\prime}-p=-\xi P+\Delta_T$.        
Acting as previously, we can derive the relation which is          
the result of independence on the choice of the gauge vector $n$.         
All the stages of derivation are the same up to the parameterization        
of the relevant matrix elements, known as Generalized Distribution Amplitudes         
(GDA) \cite{DGPT}. We are suggesting here the         
 natural set of twist-3 GDA consisting of the following         
ingredients:             
\begin{eqnarray}         
\label{par1t}         
&& \langle p, p^{\prime}| \bar\psi(0)\gamma_{\mu} \psi(\lambda n)         
|0\rangle          
=         
\int\limits_{0}^{1}d x e^{i\lambda x}         
\Biggl\{         
\tilde H_1( x; \xi, s) P_{\mu}+         
\tilde H_3( x; \xi, s)\Delta^{T}_{\mu}         
\Biggr\},         
\\         
\label{par1.1n}         
&& \langle p, p^{\prime}|\bar\psi(0)\gamma_5\gamma_{\mu} \psi(\lambda n)          
|0\rangle         
=         
\int\limits_{0}^{1}d x e^{i\lambda x}         
i\tilde H_A( x; \xi, s)         
\varepsilon_{\mu\Delta^{T} P n},         
\\         
\label{par1g}         
&& \langle p, p^{\prime}| \bar\psi(0)\gamma_{\mu}          
i\stackrel{ \leftrightarrow }         
{ \partial_{\rho}^T } \psi(\lambda n)         
|0\rangle          
=         
\int\limits_{0}^{1}d x e^{i\lambda x}         
\tilde H_1^T( x; \xi, s) P_{\mu}\Delta^{T}_{\rho}         
\\         
\label{par2g}         
&&\langle p, p^{\prime}| \bar\psi(0)\gamma_5 \gamma_{\mu}          
i\stackrel{ \leftrightarrow }         
{ \partial_{\rho}^T } \psi(\lambda n)         
|0\rangle          
=         
\int\limits_{0}^{1}d x e^{i\lambda x}         
i\tilde H_A^T( x; \xi, s) P_{\mu}\varepsilon_{\rho\Delta^{T} P n}         
\\         
\label{par1.2g}         
&&\langle p, p^{\prime} |         
\bar\psi(0)\gamma_{\mu}g A_{\rho}^T(\lambda_2 n)          
\psi(\lambda_1 n) |0 \rangle         
=         
\nonumber\\         
&&\int\limits_{0}^{1}dx_1dx_2 e^{ i x_1\lambda_1 +         
i(x_2-x_1)\lambda_2}         
\tilde B(x_1,x_2)P_{\mu}\Delta^{T}_{\rho},         
\\         
\label{par1.3g}         
&&\langle p, p^{\prime} |         
\bar\psi(0)\gamma_5\gamma_{\mu} g A_{\rho}^T(\lambda_2 n)          
\psi(\lambda_1 n) |0 \rangle         
=         
\nonumber\\         
&&\int\limits_{0}^{1}dx_1dx_2 e^{ i x_1\lambda_1 +         
i(x_2-x_1)\lambda_2}         
i \tilde D(x_1,x_2)P_{\mu}\varepsilon_{\rho\Delta^{T}Pn}.         
\end{eqnarray}         
In terms of these parameterizations, the          
$n$-independence condition for a given case reads as          
\begin{eqnarray}         
\label{n-indep_con3_gg}         
&&\frac{\partial \tilde H_1^T(x,\xi)}{\partial x}=         
\frac{\partial \tilde H_1(x,\xi)}{\partial\xi}+         
\tilde H_3(x,\xi),         
\nonumber\\         
&&\frac{\partial \tilde H_A^T(x,\xi)}{\partial x}=\tilde H_A(x,\xi),         
\end{eqnarray}         
while the QCD equations of motion take the following form         
\begin{eqnarray}         
\label{em_gg}         
&&\int\limits_{0}^{1} dy \left( \tilde D^{(S)}(x,y)         
+ \delta(x-y)\tilde H_A^{T}(y) -         
\tilde B^{(A)}(x,y) \right) =         
\nonumber\\         
&&\frac{1}{2} \tilde H_3(x,\xi)+         
\left(x-\frac{1}{2}\right)\tilde H_A(x,\xi),         
\nonumber\\         
&&\int\limits_{0}^{1} dy \left( \tilde B^{(S)}(x,y)         
+ \delta(x-y)\tilde H_1^{T}(y) -         
\tilde D^{(A)}(x,y) \right) =         
\nonumber\\         
&&\left(x-\frac{1}{2}\right)          
\tilde H_3(x,\xi)+\frac{1}{2} \tilde H_A(x,\xi),         
\end{eqnarray}         
so that within the WW approximation we have:           
\begin{eqnarray}         
\label{em_ww_gg}         
&&\tilde H_A^T(x,\xi)=\frac{1}{2} \tilde H_3(x,\xi)+         
\left(x-\frac{1}{2}\right)\tilde H_A(x,\xi),         
\nonumber\\         
&&\tilde H_1^T(x,\xi)=\left(x-\frac{1}{2}\right)          
\tilde H_3(x,\xi)+\frac{1}{2} \tilde H_A(x,\xi).         
\end{eqnarray}         
Differentiating (\ref{em_ww_gg}) with respect to   $x$       
and taking into account (\ref{n-indep_con3_gg}),         
we can reduce the system of equations (\ref{n-indep_con3_gg})          
and (\ref{em_ww_gg}) to the following form         
\footnote{ $\tilde H_\pm-$ functions are defined in complete          
analogy with $H_\pm-$ functions, see (\ref{def_pm})}          
\begin{eqnarray}         
\label{diff_sys_gg}         
&&\frac{\partial \tilde H_+(x,\xi)}{\partial x} =         
\frac{1}{x}\frac{\partial \tilde H_1(x,\xi)}{\partial\xi} ,         
\nonumber\\         
&&\frac{\partial \tilde H_-(x,\xi)}{\partial x} =         
\frac{1}{(x-1)}\frac{\partial\tilde H_1(x,\xi)}{\partial\xi} ,         
\end{eqnarray}         
         
Note that we cannot now always implement zero boundary conditions         
for both functions due to the singularities in the r.h.s.           
The only possible zero boundary conditions are:         
\begin{eqnarray}         
\label{b_con_gg}         
&&\tilde H_{+}(1,\xi)=0,         
\nonumber\\         
&&\tilde H_{-}(0,\xi)=0.         
\end{eqnarray}          
It can be easily found that the solution takes the form         
\footnote{here $\partial_\xi = \partial / \partial\xi$}         
\begin{eqnarray}         
\label{sol_gg}         
&&\tilde H_+^{WW}(x,\xi)= -         
\int\limits_{x}^{1}dy         
\frac{\partial_\xi \tilde H_1(y,\xi)}{y} ,         
\\          
\label{sol_gg2}         
&&\tilde H_-^{WW}(x,\xi)=         
\int\limits_{0}^{x}dy         
\frac{\partial_\xi \tilde H_1(y,\xi)}{y-1},          
\end{eqnarray}          
or, for           
$\tilde H_{3(A)}^{WW}-$ functions:          
\begin{eqnarray}         
\label{sol_gg.2}         
&&\tilde H_3^{WW}(x,\xi)=                   
\frac{1}{2}\int\limits_{0}^{x}dy         
\frac{\partial_\xi \tilde H_1(y,\xi)}{y-1}         
-\frac{1}{2}\int\limits_{x}^{1}dy         
\frac{\partial_\xi \tilde H_1(y,\xi)}{y} ,         
\\          
\label{sol_gg.2.2}         
&&\tilde H_A^{WW}(x,\xi)=          
-\frac{1}{2}\int\limits_{0}^{x}dy         
\frac{\partial_\xi \tilde H_1(y,\xi)}{y-1}         
-\frac{1}{2}\int\limits_{x}^{1}dy         
\frac{\partial_\xi \tilde H_1(y,\xi)}{y} .          
\end{eqnarray}         
These expressions are analogous to WW relations for the transverse polarized         
$\rho$-meson  light-cone          
distribution derived by Ball and Braun \cite{BB}.         
This is not surprising, as the momentum $\Delta_T$         
just plays the role of the polarization vector $\epsilon_T$.          
In particular, the         
structures $H_+$ and $H_-$ correspond to the  distributions with         
a definite (anti)quark spin projections\cite{BB},         
whose zero endpoint values result from their conformal properties.         
        
We can now easily obtain the decomposition the $\tilde H_{3(A)}-$functions         
on the WW type contribution and the three-particles contribution,         
as it was done above for the DVCS case. We have          
\begin{eqnarray}         
\label{full_th_3}         
\tilde H_3(x,\xi) &=&\tilde H_3^{WW}(x,\xi) -          
\frac{1}{2(1-x)} \int\limits_{0}^{1}dy         
\biggl(\tilde  B(x,y,\xi) - \tilde D(x,y,\xi)         
\biggr) +         
\nonumber\\         
&&\frac{1}{2x} \int\limits_{0}^{1}dy         
\biggl( \tilde B(x,y,\xi) + \tilde D(x,y,\xi)         
\biggr)         
\end{eqnarray}         
and          
\begin{eqnarray}         
\label{full_th_a}         
\tilde H_A(x,\xi) &=& \tilde H_A^{WW}(x,\xi) +          
\frac{1}{2(1-x)} \int\limits_{0}^{1}dy         
\biggl( \tilde B(x,y,\xi) - \tilde D(x,y,\xi)         
\biggr) +         
\nonumber\\         
&&\frac{1}{2x} \int\limits_{0}^{1}dy         
\biggl( \tilde B(x,y,\xi) + \tilde D(x,y,\xi)         
\biggr) .         
\end{eqnarray}          
In (\ref{full_th_3}) and (\ref{full_th_a}), we used the          
properties of $C$-invariance which are the following         
\begin{eqnarray}         
\label{CG}         
\tilde B(x,y,\xi) = \tilde B (y,x,\xi), \,\,         
\tilde D(x,y,\xi) =\tilde D (y,x,\xi).         
\end{eqnarray}         
         
We here stress again that          
in both cases, {\it i.e.} in the genuine expressions          
(\ref{full_th_3}), (\ref{full_th_a}) and WW approximation         
(\ref{sol_gg.2}), (\ref{sol_gg.2.2}),         
the functions $\tilde H_+$ and $\tilde H_{-}$ have, generally speaking,          
singularities          
at the points $x=0$ and $x=1$, respectively.         
However, if one assumes zero boundary condition for function $\tilde H_1$,         
required by the factorization at the leading twist level.        
these singularities are transformed to the non-zero          
finite limits at these points. They are the counterparts        
of the $x=\pm \xi$ singularities in the DVCS case.          
Anyway, these finite values          
would be sufficient to spoil factorization after the contraction          
with the short-distance subprocess cross-section.          
As in the case DVCS process, this effect can not affect         
the amplitude contracted with the polarization vector          
of the real photon.           
        
Indeed, we can derive the expression for the          
$\gamma\gamma^*\to\pi\pi$ amplitude using the developed approach         
\cite{AN00}. It reads          
\begin{eqnarray}         
\label{amp_gg}         
T_{\mu\nu}^{\gamma\gamma^*} =         
-\frac{1}{q^2}\int\limits_{0}^{1}         
\frac{dx}{x(1-x)}         
{\cal T}_{\mu\nu}^{\gamma\gamma^*},         
\nonumber\\         
\end{eqnarray}         
where         
\begin{eqnarray}         
&&{\cal T}_{\mu\nu}^{\gamma\gamma^*} =         
-q^2g_{\mu\nu}^T \biggl( 2x-1\biggr) \tilde H_1( x,\xi)          
+         
\nonumber\\         
&&\Delta_{\nu}^T \left( P+q^{\prime}\right)_{\mu}         
\biggl(         
(2x-1)\tilde H_3(x,\xi)+\tilde H_A(x,\xi)         
\biggr)+         
\nonumber\\         
&&\Delta_{\mu}^Tq^{\prime}_{\nu}         
\biggl(         
(2x-1)\tilde H_3(x,\xi)-\tilde H_A(x,\xi)         
\biggr).         
\nonumber         
\end{eqnarray}         
Note that, contrary to DVCS amplitude case, the contribution        
of $H_1$ is gauge invariant by itself. This is due to the fact, that        
parameterization of (anti)quark momenta in terms of the vector $P$ makes       
both of them collinear and independent on $\Delta_T$,        
so that one is dealing with the physical on-shell amplitude,       
which is gauge invariant.        
At the same time, only the total contributions of new quark and quark-gluon        
matrix elements is gauge invariant provided the equations of motion are        
taken into account,          
         
The contribution for the longitudinally polarized          
photon arises from the term         
\begin{eqnarray}         
\Delta_{\nu}^T \left( P+q^{\prime}\right)_{\mu}         
\biggl(         
(2x-1)\tilde H_3(x,\xi)+\tilde H_A(x,\xi)         
\biggr),         
\end{eqnarray}         
which is described by  the $\tilde H_--$structure at $x=0$ and          
by $\tilde H_+-$structure at $x=1$.         
Let us recall that at these points $\tilde H_-$ and          
$\tilde H_+$ functions are finite.         
The respective contribution to the amplitude reads as         
\begin{eqnarray}         
\label{amp_gg1}         
T_{\mu\nu}^{(1)\gamma\gamma^*} =         
-\frac{1}{q^2}\int\limits_{0}^{1}         
\frac{dx}{x(1-x)}         
\Delta_{\nu}^T \left( P+q^{\prime}\right)_{\mu}         
\biggl(         
(2x-1)\tilde H_3(x,\xi)+\tilde H_A(x,\xi)         
\biggr).         
\end{eqnarray}         
Using  (\ref{sol_gg.2}) and (\ref{sol_gg.2.2}),          
we substitute the          
$\tilde H_3^{WW}-$ and $\tilde H_A^{WW}-$functions in          
Eq. (\ref{amp_gg1}) which thereafter can be rewritten in the form          
\begin{eqnarray}         
\label{amp_gg2}         
T_{\mu\nu}^{(1)\gamma\gamma^*} =         
- \frac{\Delta_{\nu}^T \left( P+q^{\prime}\right)_{\mu}}{Q^2}         
\partial_{\xi}\int\limits_{0}^{1}         
dx \biggl(         
\frac{J_2(x,\xi)}{1-x} + \frac{J_1(x,\xi)}{x}         
\biggr),         
\end{eqnarray}          
where          
\begin{eqnarray}         
J_1(x,\xi) = \int\limits_{0}^{x}dy         
\frac{\tilde H_1(y,\xi)}{y-1} ,\quad         
J_2(x,\xi) = \int\limits_{x}^{1}dy         
\frac{\tilde H_1(y,\xi)}{y}.         
\end{eqnarray}         
After simple manipulations, we derive the following expression         
\begin{eqnarray}         
\label{amp_gg3}         
T_{\mu\nu}^{(1)\gamma\gamma^*} =         
\frac{\Delta_{\nu}^T \left( P+q^{\prime}\right)_{\mu}}{Q^2}         
\int\limits_{0}^{1}         
dy \partial_{\xi} \tilde H_1(y,\xi)         
\biggl(         
\frac{\log{(1-y)}}{y} - \frac{\log{y}}{1-y}         
\biggr),         
\end{eqnarray}         
which coincides with the expression in \cite{K&M}.          
At the same time, the genuine twist-3 contribution is new.          
         
\noindent         
{\it Conclusions.}   
In this paper we have presented the new         
derivation of the WW relations between the parameterizing         
twist-2 and kinematical twist-3 functions.         
This simple approach is based on the use of          
the equations of motion and on the requirement of          
independence of          
the amplitudes from the gauge-fixing light-cone vector.            
        
We have found the explicit          
expression for the genuine twist-3 functions for DVCS and          
$\gamma\gamma^*\to 2\pi$ processes.         
We also have computed the amplitude of           
$\gamma^{\star}\gamma\to\pi\pi$ including         
complete twist-3 corrections.         
        
The counterparts of the singularities appearing in the        
DVCS case are provided by the finite values of generalized         
distribution amplitudes in the endpoints, which are         
related to their conformal properties. Therefore, one may hope        
that appearance of the singularities, as well as their         
cancellation in the physical amplitudes, have some general reasons.         
        
We have also shown that the contributions from the three-particles          
correlators posses the same singularities as          
the WW parts which also do not contribute to the          
physical amplitudes at the twist-3 level.

\noindent         
{\it Acknowledgments.}    
We are thankful to V.~Braun, D.~Ivanov, N.~Kivel,         
B.~Pire, A.~Sch\"{a}fer and C.~Weiss for discussions.         
This work has been supported in part by EC-RTN Network         
ESOP, Contract HPRN-CT-2000-00130, by DGESIC (Spain) under         
contract PB97-1401-C02-01, by DFG and by RFFI Grant 00-02-16696.

\end{document}